\documentclass{iopart}
\usepackage{graphicx}
\usepackage{natbib}
\usepackage{url}

\def\be{\begin{equation}}
\def\ee{\end{equation}}

\begin{document}

\title[SMBBH search]{A hierarchical search for gravitational waves 
from supermassive black hole binary mergers}

\author{I.~W.~Harry, S.~Fairhurst, B.~S.~Sathyaprakash}

\address{Cardiff School of Physics and Astronomy,
Cardiff University, Queens Buildings, The Parade, Cardiff. CF24 3AA
}

\eads{\mailto{ian.harry@astro.cf.ac.uk}
\mailto{Stephen.Fairhurst@astro.cf.ac.uk}
\mailto{B.Sathyaprakash@astro.cf.ac.uk}} 

\begin{abstract}

We present a method to search for gravitational waves from coalescing
supermassive binary black holes in LISA data.  The search utilizes the
$\mathcal{F}$-statistic to maximize over, and determine the values of,
the extrinsic parameters of the binary system.  The intrinsic parameters
are searched over hierarchically using stochastically generated
multi-dimensional template banks to recover the masses, sky location and
coalescence time of the binary.  We present the results of this method
applied to the mock LISA data Challenge 1B data set.

\end{abstract}

\section{Introduction}
\label{sec:intro}

There is growing evidence that some fraction of quasars \cite{Nature:0804},
and X-ray and infrared sources \cite{KomossaEtAl} host supermassive binary 
black holes (SMBBH) that are potential sources of gravitational 
radiation.  The late time evolution of such systems is dominated 
by the emission of gravitational waves, the radiation back reaction
torque driving the system to coalesce. The Laser Interferometer 
Space Antenna (LISA) \cite{Bender95} targets gravitational waves from these
systems in the frequency range of $[{\rm few}\,\times 10^{-5},\, 0.1]$ Hz
which corresponds to SMBBH of masses in the range 
$[10^4,\, {\rm few}\, \times 10^7]M_\odot$. 
Within a redshift of $z\sim 10,$ SMBBH coalescence rates 
could be as high as several tens per year but depending 
on the way galaxies and black holes at their cores formed 
the rates could be several hundreds per year \cite{Rees:2007nc,Sesana:2007sh}.
Even at redshifts $z\sim 10$ mergers can be detected by
LISA, expected amplitude signal-to-noise ratio (SNR) being in excess
of several thousands for sources at a red-shift of $z=1,$ implying
an SNR of $\sim 10$ even at redshifts of $z \simeq 10.$   LISA is, 
therefore, an excellent probe of the seed black holes that are 
believed to be responsible for the formation and evolution of 
galaxies \cite{Rees:2007nc,Sesana:2007sh,Volonteri:2002vz}
and the large-scale structure in the Universe and
it is important to be able to detect SMBBH mergers at as low
an SNR as $\sim 10.$

SMBBH mergers from redshifts up to about $z\sim 3$ can be detected
without any sophisticated data analysis, although accurate models 
of the merger dynamics would be needed for parameter extraction. 
Indeed, these sources will be so bright that one has to worry 
about systematics due to our limited theoretical understanding of 
their dynamics \cite{Cutler:2007mi}. At larger red-shifts, however,
it would be necessary to employ data analysis techniques that
are sensitive to weaker signals. 
This is an important goal for LISA as there is significant 
uncertainty in when the first seed black holes and galaxies might 
have formed and it would be good to be able to probe as far back as
a red-shift of $z \sim 10$-$15.$ At a red-shift of $z,$ an SMBBH of intrinsic
total mass $M$ would appear in LISA to have a red-shifted total mass
of $(1+z) M.$ Thus, at $z=10$ LISA would probe masses that are 
intrinsically 11 times smaller than at $z=0.$  Therefore, searching
for SMBBH at higher red-shifts would probe smaller masses too.

In addition to SMBBH mergers LISA will observe a host of other sources
(see \cite{LISA:MSO2007} and references therein).  These include binary
white dwarfs in the Milky-way (both a stochastic signal from an
unresolved background population and continuous signals from resolved
foreground sources), inspirals of small black holes into supermassive
black holes (again a stochastic background from overlapping sources and
a foreground of individual sources), etc. Analysing LISA data and
resolving tens of thousands of signals belonging to different classes is
unprecedented and likely to be a daunting task.  Matched filtering is a
very powerful approach that has been successfully used in several
applications to dig weak signals from noisy backgrounds. For example,
matched filtering using a bank of templates has been extensively used in
searching for gravitational wave signals in ground based detectors, see
for example \cite{Abbott:2007xi, Abbott:2008uq}. In this paper we report
the results from a hierarchical matched filtering algorithm to search
for SMBBH mergers. 

From a computational point of view, however, matched filtering is very
expensive, computational cost increasing as a power-law of the number of
search parameters.  While the parameter space of stellar mass binaries
consisting of non-spinning black holes is only two-dimensional, the
number of parameters in the case of SMBBH, even while neglecting spins,
is quite large. This is because the source's position relative to LISA
changes during the course of observation, causing a modulation in the
signal's amplitude and phase that must be taken into account in the
search templates as well as the waveform's polarization angle.  Thus,
the computational cost of a naive implementation of a matched filtered
search would be formidable. For example if we were to conduct a one
stage match-filtered search on the MLDC 1B dataset to get parameter accuracies
to similar levels to those we quote in our results section we estimate
that you would require $10^{19}$ templates. Of course \textit{detecting} 
a signal as
loud as the ones used in the MLDC requires significantly less templates, and
can be done by placing templates in mass space only.
We therefore developed a hierarchical approach in
which the goal was to zoom-in onto an interesting region of the
parameter space in several steps, each of which uses a progressively
greater density of templates. We tested our algorithm on the {\em
training} and {\em challenge} data sets from Challenge1B of the Mock
LISA Data Challenge (MLDC).%
\footnote{The LISA International Science Team
has put together a task force to develop a set of data analysis
challenges \cite{Arnaud:2007vr} of ever increasing complexity
\cite{Arnaud:2007jy} to encourage data analysts to explore and test
their search algorithms on simulated data.  The most recent release of
challenge data sets was Challenge 1B, a rerun of the Challenge 1,
consisting of the simplest possible data sets with only one inherent
signal.} 

In the SMBBH coalescence section of the 
data set the challenge was to detect and characterize one 
SMBBH coalescence buried in LISA instrumental noise only. 
Two datasets were released, one where the coalescence was in 
the middle of the observation period and a second where the 
coalescence was two months after the observation period ended. 
We only took part in the Challenge where the binary coalesced 
during the observation time.  

For our templates we used post-Newtonian waveforms at the
second post-Newtonian order. We tapered the end point of
our templates to prevent the bleeding of spurious power in 
the Fourier domain arising from the step function that is 
implicit if the waveform were to be terminated abruptly. 
The signal is characterized by nine independent parameters. 
We separate these into the `intrinsic' parameters consisting of 
the two component masses, the binary's position on the sky and its
epoch of coalescence and `extrinsic' parameters comprising
the inclination angle, the polarization phase, the coalescence 
phase and the distance to the binary. We devised a search that
was capable of determining all these parameters in an efficient
manner, albeit not to accuracies that are theoretically possible.
Let us note, however, that the goal of this exercise is not
to measure the parameters accurately but to efficiently 
detect the signal and constrain the parameter space well enough
so that other techniques, such as the Markov Chain Monte Carlo, 
can be deployed in a follow-up study to determine the parameters 
more accurately.

Other groups have, of course, participated in the search for SMBBHs in
the Mock LISA Data Challenges, and their methods differ from ours
\cite{Arnaud:2007vr, Babak:2007zd, mldc1b, Cornish:MLDC1, Brown:2007se,
Camp:MLDC1, Cornish:MLDC2}. The majority of these searches involve a
variety of methods to detect the source and constrain somewhat its
parameters followed by a Markov Chain Monte Carlo followup to determine
accurately all the parameters.

\section{Search Method}
\label{sec:method}

A search for supermassive binary black hole signals in the LISA data
requires, in general, the determination of seventeen parameters.  In this
paper, and in Challenge 1B of the MLDC, spins of the component black holes
are ignored, restricting to non-spinning components.  In addition, 
the orbit is assumed to be circularized sufficiently by the time it
enters LISA's sensitive band that eccentricity can be ignored.  
This allows us to neglect eight parameters leaving the parameters 
of interest to be:
\begin{itemize}
\item The masses of the two components of the binary, $M_{1}$ and
$M_{2}$.  It is often convenient to express the mass parameters in terms
of the chirp mass $\mathcal{M}$ and reduced mass $\mu$, defined
as
\begin{equation}
  \mathcal{M} = (M_{1} + M_{2})^{2/3} \, \mu^{3/5} 
  \quad \mathrm{and} \quad
  \mu = \frac{M_{1} \, M_{2}}{M_{1} + M_{2}} \, .
\end{equation}
\item The time that the binary coalescences, $t_{c}$, which is assumed to
be within the LISA data.
\item The sky location of the binary, determined by its ecliptic
latitude, $\beta$, and longitude, $\lambda$.
\item The orientation of the binary system, given by the inclination
angle, $\iota$, and polarization angle, $\psi$.
\item The initial phase of the binary, $\varphi_{o}$.
\item The luminosity distance to the binary, $D_{L}.$
\end{itemize}

In this search, we make use of the `$\mathcal{F}$-statistic'
\cite{fstatpaper} to analytically maximize over four of the parameters
introduced above (the `extrinsic parameters'): the distance to the
binary, and its inclination, polarization and initial phase.  This
procedure is discussed in Section \ref{sec:fstat}.  The remaining five
parameters are determined by searching over stochastically generated
template banks \cite{stochasticbank, randbankpaper}.  Since some
parameters, in particular the chirp mass and coalescence time, are more
easily determined we employ a hierarchical search whereby we obtain good
estimates of these parameters before refining our search to determine
the full parameter set.  The stochastic bank is described in Section
\ref{sec:randombank} and the hierarchical search method is discussed in
\ref{sec:search}.  Results of this search applied to the MLDC 1B data
are presented in Section \ref{sec:results}.

\subsection{Matched Filtering with the F-statistic}
\label{sec:fstat}

When the signal waveform is well known, the technique of
matched-filtering is typically used to search for the signal (see, for
example \cite{wainstein:1962, FinnCh93}).  The $\mathcal{F}$-statistic
is an elegant way to maximize over the extrinsic parameters, and thereby
simplify the search. It is used in a number of other searches by other
groups for the MLDC and in the pulsar search with ground-based detectors
(for example \cite{Abbott:2008uq, Whelan:2008yb}. We quickly recap how
it is used in this section.  We follow closely Ref.~\cite{fstatpaper}
when describing this method. Given a template waveform $h$ and the data
$s$, we calculate the {\em likelihood function}, defined as:
\begin{equation}\label{eq:likelihood} 
 \ln \Lambda = \left\langle s , h \right\rangle 
  - \frac{1}{2}\left\langle h , h \right\rangle \, ,
\end{equation}

where the inner product between the signal and template is given by 
\begin{equation}\label{eq:inner_product} 
  \langle s, h \rangle = 4 \, \mathrm{Re} 
  \left[\int\limits_0^{\infty} \frac{\tilde{h}^{\ast}(f) 
  \tilde{s}(f)}{S_h(f)}df\right] \, ,
\end{equation}
and $S_h(f)$ is the power spectral density of the LISA detector.

For a signal present in the LISA data, it can be shown that the
gravitational wave signal can be decomposed as \cite{Krolak:2004xp}
\begin{equation}
  h(t) = \sum^4_{i=1} A_{i}(D_{L}, \varphi_{o},\iota,\psi) \cdot 
  h_{i}(t; \, t_{c}, M_{1}, M_{2}, \lambda, \beta) \, . 
\end{equation}
The amplitudes $A_{i}$ are functions only of the extrinsic
parameters: $D_{L}, \varphi_{o}, \iota$ and $\psi$. The $h_{i}(t)$ are
functions of the remaining, intrinsic, parameters only. The benefit of
expressing the waveform in this manner is that it is straightforward to
maximize the likelihood parameters over these $A_{i}$ by requiring
\begin{equation} 
  \frac{\partial \ln \Lambda}{\partial A_i} = 0 \, .
\end{equation}
With a bit of algebra this can be shown to be equivalent to
\begin{equation}\label{eq:ai_values}
  A_{i} = \sum^4_{j=1} \mathcal{M}_{ij}^{-1} \left\langle s , h_{j}
  \right\rangle 
  \quad \mathrm{where} \quad 
  \mathcal{M}_{ij} = \left\langle h_i , h_j \right\rangle \, . 
\end{equation}

Therefore, the $A_{i}$ can be determined from the $\left\langle h_i ,
h_j \right\rangle$ and $\left\langle s , h_i \right\rangle$.
Furthermore, for each possible set of values for $A_{i}$ we obtain a
unique value for the four extrinsic parameters: distance $D_{L}$,
initial phase $\varphi_{o}$, inclination angle $\iota$ and polarization
angle $\psi$.  However, there remain implicit degeneracies in these
values. Specifically, as we use only the dominant, $2\Phi$ harmonic in
the waveform, there is a degeneracy in the initial phase corresponding
to $\varphi_{o} \rightarrow \varphi_{o} + \pi$.  The same degeneracy
exists for the polarization angle.  Additionally, a system with
polarization $\psi$ and phase $\varphi_{o}$ is indistinguishable from
one with values $\psi + \frac{\pi}{2}$ and phase $\varphi_{o} +
\frac{\pi}{2}$.  Finally, by substituting the expression for $A_{i}$
from (\ref{eq:ai_values}), the likelihood expression becomes

\begin{equation}\label{eq:maximized_likelihood}
  \ln \Lambda = \frac{1}{2} \sum^4_{i,j=1} 
  \left\langle s , h_{i} \right\rangle 
  \mathcal{M}_{ij}^{-1} 
  \left\langle s , h_{j} \right\rangle \, .
\end{equation}

In the above discussion, we have used the gravitational wave strain $h(t)$
in discussing the $\mathcal{F}$-statistic.  In the MLDC, the signals
were released in the form of time delay interferometry (TDI)
variables $X$,$Y$ and $Z$ \cite{Tinto:2002de, Vallisneri:2005ji}.  These
TDI variables are used as a way of cancelling the laser phase noise in
the output of LISA.%
\footnote{It is assumed for the purposes of this Challenge that this
time delay interferometry process will completely cancel all of the
laser phase noise.}  
The $\mathcal{F}$-statistic method is equally applicable to the TDI
variables.  To maximize the efficiency of our search method we
simultaneously utilize two of the TDI outputs, $X$ and $Y$,
to conduct our search.  We do not use the $Z$ output since the gravitational
wave content in it can be constructed from the other two and is
therefore not independent.

It is a trivial matter to convert the one-detector search outlined above
to a two-detector TDI search.  We simply rewrite our likelihood function
as:
\begin{equation}
  \ln \Lambda = \left\langle s_X , h_X \right\rangle  
  + \left\langle s_Y , h_Y \right\rangle
  - \frac{1}{2}\left\langle h_X , h_X \right\rangle 
  - \frac{1}{2}\left\langle h_Y , h_Y \right\rangle \, , 
\end{equation}
where the subscripts $X$ and $Y$ denote the data or template appropriate
for either the $X$ or $Y$ TDI data stream. 
\footnote{Strictly speaking this expression is incorrect for the $X$ and
$Y$ channels as the noise in them is correlated.  It is, of course,
preferable to use the synthetic $A$ and $E$ variables which are
generated from $X$, $Y$, $Z$ and are independent.  Due to time
constraints, for this challenge we did not get around to moving the code
over to $A$ $E$ and $T$.  This has since been implemented.}
The $\mathcal{F}$-statistic maximization can similarly be extended to
the two detector search.  In this case, the expressions in
(\ref{eq:ai_values}) and (\ref{eq:maximized_likelihood}) generalize to
include a summation over detector.  For example:
\begin{equation}
  \mathcal{M}_{ij} = \left\langle h_{i,X}, h_{j,Y} \right \rangle 
  + \left\langle h_{i,Y}, h_{j,Y} \right \rangle \, . 
\end{equation}

\subsection{Stochastically generated template bank}
\label{sec:randombank}

Even after maximizing over the `extrinsic' variables, there are still
five remaining, `intrinsic' parameters that we would like to determine.
We utilize a template bank to search over this five dimensional
parameter space \cite{Owen96, OwenSathyaprakash98}.  Existing templated
searches for gravitational waves from binary coalescences in  ground-based 
detectors utilize a two, or at most three, dimensional parameter
space.  Geometrical placement algorithms exist \cite{Babak:2006ty,
Cokelaer:2007kx} to deal with the problem of efficiently placing
templates in these parameter spaces. However, when the parameter space
becomes higher dimensional we have two problems with using these
geometric placement methods. Firstly, there is no known optimal
placement algorithm for dimensions higher than two. It has been shown
\cite{Prix:banks} that placing a square lattice of templates becomes
grossly inefficient in higher dimensions. Additionally, when the signal
manifold becomes curved it is unclear how to construct these geometrical
lattices. The signal manifold for the SMBBH search in LISA data suffers
from both of these issues.  Therefore, we use the method outlined in
\cite{stochasticbank, randbankpaper} and create {\em stochastically}
generated template banks. Other implementations of randomly generated template
banks can be found in \cite{messenger,Babak:randbank}.

The final stochastic bank is designed so that for any signal in the
parameter space, at least a fraction $M$ of the potential signal power is
recovered.  This is most easily understood by introducing the notion of
overlap between two templates as:
\begin{equation}\label{eq:overlap}
  \mathcal{O}(h_{1}, h_{2}) = \frac{\left\langle h_{1} , h_{2} 
  \right\rangle}{|h_{1}| |h_{2}|} \, ,
\end{equation}
where the norm of the template is defined as $|h|^{2} = \langle h , h
\rangle$.  When the two templates are identical, the overlap will be
unity.  Then, given any signal $s$ in the parameter space, we require
that 

\begin{equation}\label{eq:match}
  \mathrm{Max}_{I}\left( \mathcal{O}(s, h_{I}) \right)
  \ge M \, 
\end{equation} 
where $I$ labels the templates in the bank.  The parameter $M$ is known
as the minimal match. 

\begin{figure}[htp]
  \centering
  \begin{minipage}[t]{0.45\linewidth}
    \centering
    \includegraphics[width=\linewidth]{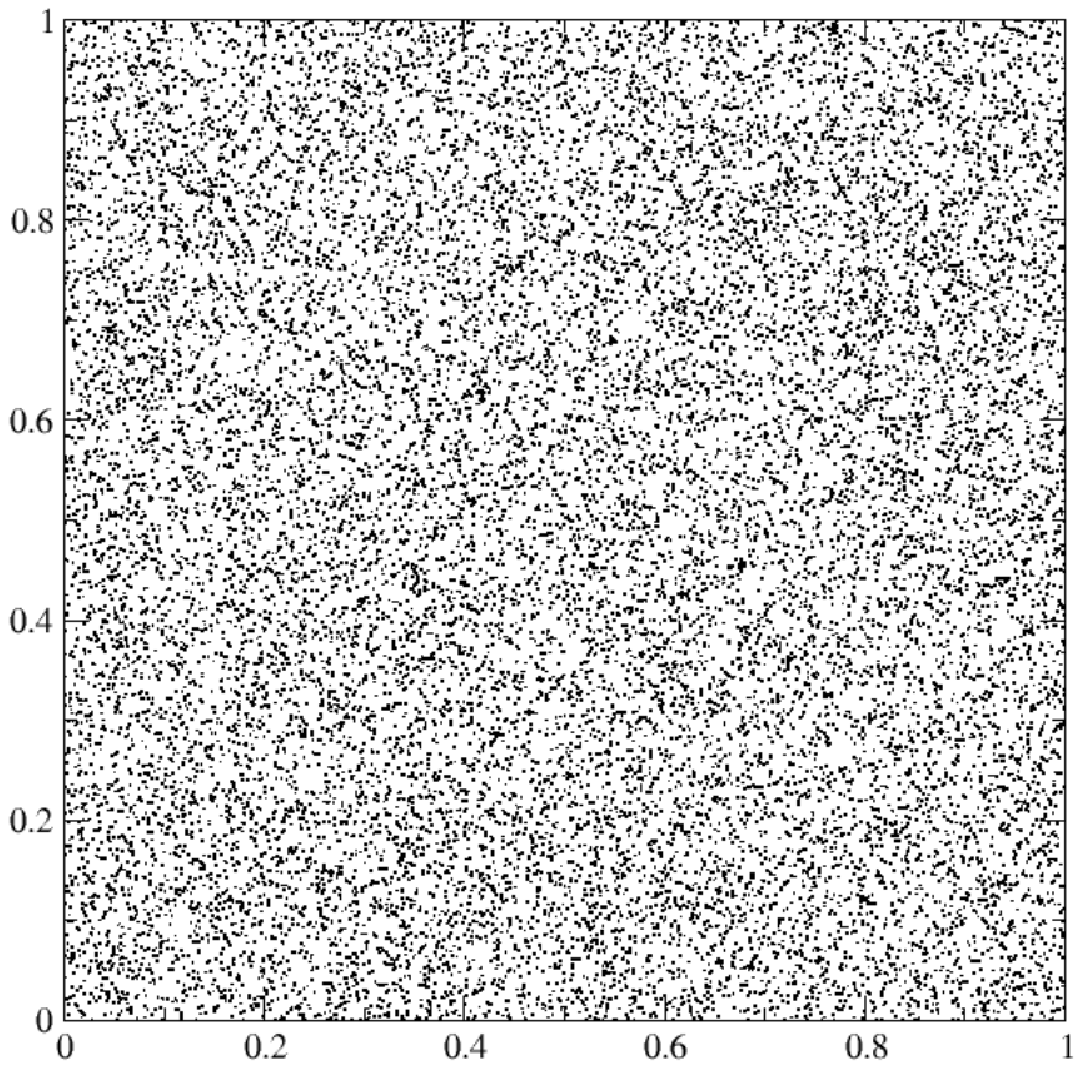}
  \end{minipage}
  \hspace{0.05\linewidth}
  \begin{minipage}[t]{0.45\linewidth}
    \centering
    \includegraphics[width=\linewidth]{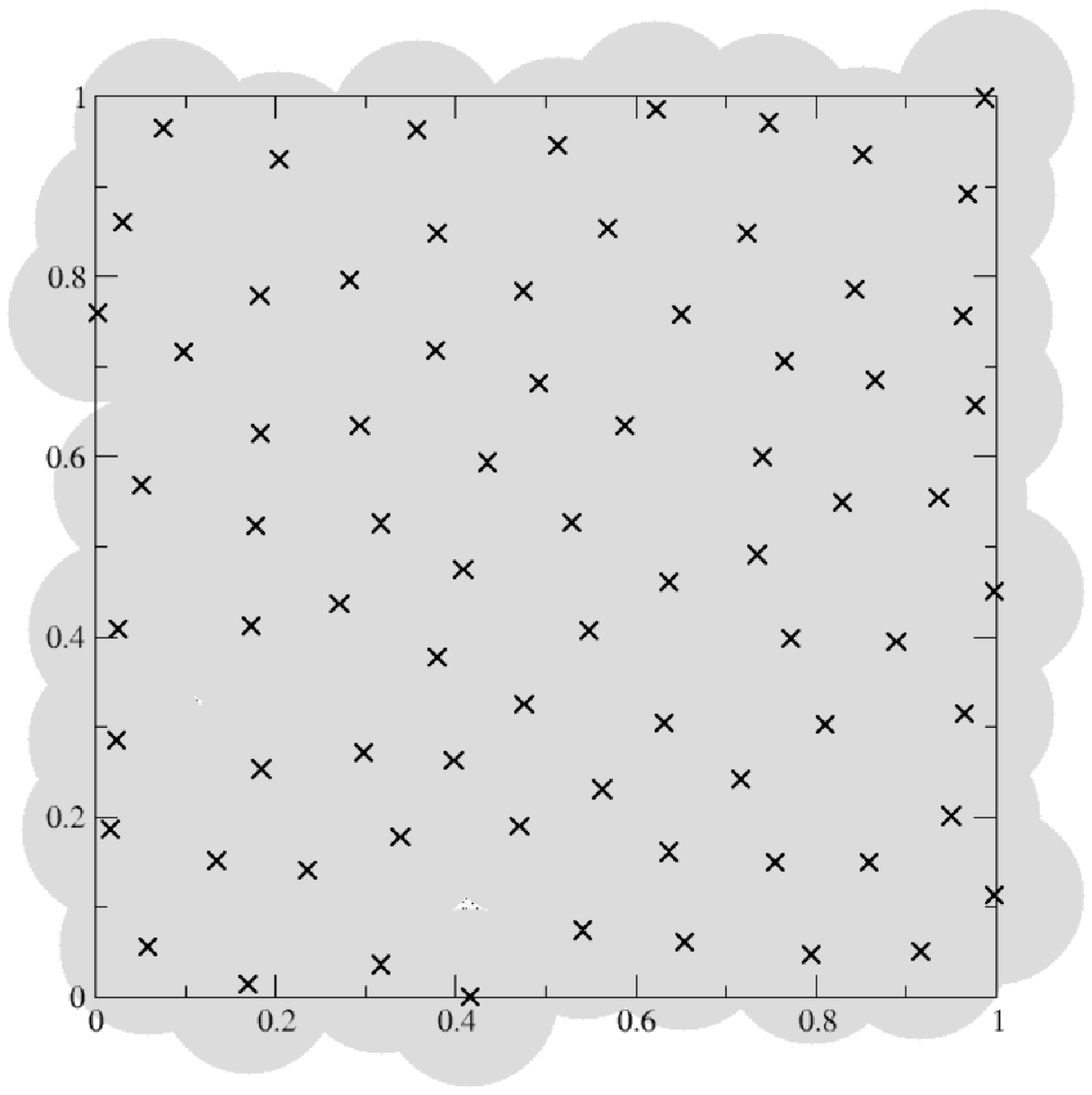}
  \end{minipage}
  \caption{\label{fig:randbank} 
The left hand figure shows an initial set of over-dense
templates which are placed stochastically in an arbitrary flat parameter space
(where the parameter space metric is the identity). These are subequently 
filtered to reduce the number, demanding a maximum overlap between templates
of 0.995.  The right hand figure shows the final
template bank generated by the stochastic bank placement procedure.  The
shaded circles surrounding each template are the regions of parameter
space which are covered by the template.  As can be seen, the vast
majority of the space is adequately covered. The axes here are arbitrary.}
\end{figure}

We begin by choosing a randomly generated set of densely spaced points
in the parameter space that are the candidate templates, see
Fig.~\ref{fig:randbank}.  There are significantly more initial points
than would be needed to cover the space with an appropriate density of
templates to satisfy (\ref{eq:match}).  We subsequently go through the
candidate templates and remove those which are superfluous. 
 
More specifically, we select an initial template and remove any other
templates from the bank which are redundant as they are too close to the
initial template.  This is simplified by calculating the metric on the
parameter space,
\begin{equation}\label{eq:metric}
  g_{ij}(\theta) = \left\langle \frac{\partial \hat{h}}{\partial \theta_{i}} , 
  \frac{\partial {\hat h}}{\partial \theta_{j}} \right\rangle, \quad 
\hat h = \frac{h}{|h|},
\end{equation}
at the location of the initial template (whose parameters we denote by
$\theta$).  Then, we can approximate the overlap between this template
and any other in the bank as 
\begin{equation}
  \mathcal{O}(h(\theta), h(\theta + \delta \theta)) \approx
  1 - \frac{1}{2}g_{ij} (\delta \theta)^{i} (\delta \theta)^{j} \, .
\end{equation}
If the overlap is greater than the required minimum, the second template
is discarded.  Having tested every template, we then move on to one of
the surviving templates and repeat the proceduce, again discarding
templates which are too close to the selected template.  This process is
repeated until all templates have been tested.  The method is efficient
since the costly process of computing the metric is only performed at the
location of surviving templates, not at the location of all initial
templates.  

The final template bank will cover the majority of the space to the
desired accuracy.  However, since the initial process of placing points
is stochastic, we cannot guarantee that the entire space will
necessarily be covered appropriately. To try to be sure that the parameter
space is adequately covered we choose a number of initial seed points, which
we believe would be enough to cover the parameter space and after the bank
has been generated we generate a further 20,000 seed points and test their
overlap with the rest of the bank. The percentage of these seed points that
have an overlap with the template bank of more than the required minimum can
then be used as a measure of how well the parameter space is covered. If the
coverage is not sufficient we can generate more seed points and extend the
template bank. An alternative method would be to, instead of using a 
predetermined number of initial seed points, run the template bank generation
until a specfic number of seed points have been rejected concurrently, we 
would then consider this template bank adequate \cite{stochasticbank}.
Fig.~\ref{fig:randbank} shows an
example template bank that results from this procedure. Using this
method we are left with a stochastically generated template bank that is
capable of covering any parameter space in any number of dimensions.

While the stochastic bank generation is generic, there are certain
subtleties which arise in employing it for the $\mathcal{F}$-statistic
search for SMBBH described in Section \ref{sec:fstat}.  First note that,
in contrast to searches for binaries in ground-based detectors, we must
include the coalescence time when generating the template bank.  Binary
coalescence signals in ground-based decectors last at most $\sim 1,000$
seconds, during which the motion of the Earth, and the detector,
can be neglected.  Hence, the waveforms of binaries with different
coalescence times differ only by a time-shift and amplitude rescaling.
However, SMBBH signals spend several months in LISA's sensitive band,
during which LISA completes a significant fraction of an orbit
around the Sun.  Consequently, the template-shape depends on the
coalescence time, and this parameter must be included in the template
bank.

Next, we consider the effect of maximization over the four extrinsic
parameters in the $\mathcal{F}$-statistic.  This is dealt with by
generating a metric on the full parameter space and projecting down to
the five-dimensional subspace (see \cite{Babak:2006ty} for details).  A
complication arises in that the projected metric depends upon the value
of three of the extrinsic parameters $\iota$, $\psi$ and $\varphi_{o}$.%
\footnote{It is immediate from the definition of the metric
(\ref{eq:metric}) that the distance, $D_{L}$, will not affect the metric
at all and can safely be neglected.}
This is a well known issue, see for example \cite{Prix:2006wm}.  To
proceed, we simply choose a fiducial value of $0.5$ radians for these
angles.  The value was chosen arbitrarily, ensuring that none of the four
$A_{i}$ values was zero and they would all contain contributions
from both gravitational-wave polarizations.  

To generate the metric, we calculated the inner product
(\ref{eq:inner_product}) for the $X$-detector using gravitational-wave
strain $h(t)$ rather than the TDI variables.  This introduces two
additional approximations.  First, by using the strain, rather than the TDI
variables, we are neglecting the directional dependence of the
detector's response function and implicitly working at the long-wavelength
approximation.  Second, we have performed the search using both the $X$ 
and $Y$ data streams while only the metric for $X$ was used to generate the
template bank.  The above simplifications will mean that the stated
minimal match of the metric would not have been achieved.  However, in performing the
search, as described in Section \ref{sec:search}, we continually
refined the template bank to determine the correct parameter values and
did not rely on the minimal match to decide stopping conditions.

\subsection{Hierarchical Search Technique}
\label{sec:search}

Populating even the reduced, 5-dimensional parameter space with
sufficient templates to determine the binary's parameters to the
required accuracy would necessitate far more templates than could feasibly
be filtered (as discussed in the introduction).  
Thus we must employ a
hierarchical method to search for the parameters. 

We began by match-filtering the data against a bank comprised of
templates that are sparsely spaced and placed in only the 
two-dimensional space of mass parameters.  This bank enabled us to make an
initial estimate of the binary's masses and coalescence
time  with 1,000 templates in the allowed
range of masses, setting the sky location arbitrarily to $\lambda =
0.5$ and $\beta = 0.5$ for all templates and fixing the coalescence time
to be the value at the beginning of the allowed range.  This enabled us to estimate the
chirp mass and reduced mass to within 30\% accuracy and coalescence time
to within 10,000 seconds.

We then placed a second bank of 1,000 templates within a reduced range of the
parameter space, using the best estimate of the coalescence time, sky
locations again set to $\lambda=\beta=0.5$ and repeat the process.  By this method we
could estimate the chirp mass to at least $\pm5\%$, the reduced mass to at
least $\pm10\%$ and the coalescence time to within 10,000~s.  Using these
initial estimates we were then able to place a template bank with
restricted parameter ranges to determine all five of the `intrinsic'
parameters. 

The final step in determining the parameters could be performed by 
two different methods. The first method involves placing a template bank over
the full five-dimensional parameter space and using a hierarchical procedure to `zoom in' on
the true values of the binary's parameters.  While this is the preferred
search method, a large number of templates are still required to fill
this reduced five-dimensional template space, to do this in one step would
require $10^{13}$ templates. We would thus have to use a heirarchical procedure
to construct a series of five dimensional banks, but, this search can still become
computationally costly.  An alternative technique is to alternate
between placing two-dimensional template banks in the mass space, using the
best current estimates of coalescence time and sky location, and placing
three-dimensional banks in sky location and coalescence time, using the
best current estimates for the masses. This method is computationally
quicker as we limit the template bank size to under 1000 templates for the
two dimensional case and under 10,000 templates for the three dimensional
banks. However, much more than when using 5 dimensional banks, care must be
taken to avoid `zooming in' on secondary maxima.  For example, LISA has
similar sensitivity to binary systems on opposite sides of the sky, so
restricting the range of sky locations used in our template bank
searches is not trivial. 

The figures quoted above for template bank size and parameter accuracy
are those for the binary systems in the MLDC datasets where the SNR is
very large (approx. 500).  For SMBBH systems where the SNR is
significantly lower the main issue would be whether any templates at the
initial stage were similar enough to the signal to pick it up. If so,
the parameter accuracies at this stage would be similar as they are
limited by the template spacing. Further investigation is warranted to
determine what strength of signals can be detected by this method, how
many more templates are needed at initial stage to detect weaker signals
and how final parameter accuracy depends on SNR.

In future searches using this method it would be desirable to automate
the heirarchical technique. To do so, we would need to quantify how
many iterations are needed to adequately determine the parameters and
how much each iteration reduces the possible range of values for each
parameter. Although this method is still under development, it is
interesting to note that it uses a comparable number of templates as the
MCMC search implemented in \cite{Cornish:MLDC2}. It is also worth noting
that in a template bank based search it is straightforward to
parallelize the search over numerous computers.

\section{Results}
\label{sec:results}

The MLDC Challenge 1B data set for SMBBH consists of one year of
simulated LISA data with a single supermassive binary black hole
coalescence occuring during the year.  In addition, a
``training'' data set was released for which the binary's parameters were
also made public.  Due to unforeseen technical issues we were unable to
run as full an analysis as we would have liked on the challenge dataset,
and our results reflect this.  Therefore, we have also included the
results from the training run, as they provide a more accurate
reflection of the sensitivity of our current search technique.  The
released training data parameters were not used in running the search,
as it was treated as a warm up to the challenge.  For both training and
challenge results we have taken into account the parameter degeneracies
discussed in Section \ref{sec:fstat} by choosing the values of
polarization and initial phase that are closest to the true values. 
\begin{table}
{\small
\begin{tabular} {l | llll }
Parameter & True Value & Our Value & Error & Fract.~Error\\
\hline
 Chirp Mass, $\mathcal{M}$ ($M_{\odot}$) &$1.3769 \times 10^{6}$ 
 &$1.3772 \times 10^{6}$ & $360$ 
 &$2.6 \times 10^{-4}$ \\
 Symmetric Mass Ratio, $\eta$ & 0.1959 & 0.1972 & 0.0013 & -- \\
 Ecliptic Latitude, $\beta$ & 1.028 & 1.072 & 0.044 & -- \\
 Ecliptic Longitude, $\lambda$ & 5.050 & 5.037 & 0.013 & -- \\
 Coalescence Time, $t_{c}$ (s) & $17523096.4$ & $17523090$ & $6.4$ & -- \\
 Polarization Angle, $\psi$ & 0.826 & 0.668 & 0.158 & -- \\
 Inclination Angle, $\iota$ & 2.846 & 2.313 & 0.533 & -- \\
 Initial Phase, $\varphi_{o}$ & 1.844 & 1.836 & 0.048 & -- \\
 Luminosity Distance, $D_{L}$ (Gpc) & $36.3$ & $26.6$ & 
 $9.6$ & 0.27 \\

\end{tabular}
}
\caption{Table showing the results of our analysis on a training dataset.}
\label {tab:training}
\end{table}

\begin{table}
{\small
\begin{tabular} {l | llll }
Parameter & True Value & Our Value & Error & Fract.~Error \\
\hline
 Chirp Mass, $\mathcal{M}$ ($M_{\odot}$) & $2.6832 \times 10^{6}$ & 
 $2.6904 \times 10^{6}$ & $7178.8$ & $2.68\times10^{-3}$ \\
 Symmetric Mass Ratio, $\eta$ & 0.2159 & 0.2316 & 0.0158 & -- \\
 Ecliptic Latitude, $\beta$ & 1.139 & -0.235 & 1.374 & -- \\
 Ecliptic Longitude, $\lambda$ & 3.931 & 3.382 & 0.549 & -- \\
 Coalescence Time, $t_{c}$ (s) & $15045887.8$ & $15046429.6$ & $541.2$ &
-- \\
 Polarization Angle, $\psi$& 6.063 & 5.941 & 0.123 & -- \\
 Inclination Angle, $\iota$ & 1.939 & 1.252 & 0.687 & -- \\
 Initial Phase, $\varphi_{o}$ & 0.213 & 1.031 & 0.818 & -- \\
 Luminosity Distance, $D_{L}$ (Gpc) & $10.7$ & 
 $26.0$ & $15.3$ & 1.43 \\
\end{tabular}
}
\caption{Table showing the results of our analysis on the official
challenge dataset.}
\label {tab:challenge}
\end{table}

The results from the training data set are presented in Table
\ref{tab:training}, while the Challenge results are shown in Table
\ref{tab:challenge}.  It is interesting to compare our results to those
obtained by other groups applying different methods to search for SMBBH
coalescences in the Mock LISA Data Challenge \cite{Arnaud:2007vr,
Babak:2007zd, mldc1b, Cornish:MLDC1, Brown:2007se, Camp:MLDC1,
Cornish:MLDC2}.  It is clear that our Challenge results are
substantially less accurate, for reasons described above.  However, our
results from the training data set are comparable to those obtained
using other methods.  In particular, it is gratifying to see that we
were able to obtain the correct sky location.  Furthermore, the sky
location is recovered to within a few square degrees, which is the
accuracy required to make an optical followup feasible (see for example
\cite{Tyson:2003kb}).

\section{Summary and Future Plans}
\label{sec:summary} 

We have presented a hierarchical, template-based search method for SMBBH
in LISA data.  This method makes use of the $\mathcal{F}$-statistic to
reduce the parameter space for non-spinning black holes from nine to
five dimensions, and then employs a stochastically generated template
bank to search over the remaining parameter space.  This method has been
applied to perform a search on the data from Challenge 1B of the MLDC.
We were able to successfully locate the signal and, in the case of the
training data, recover its parameters with good accuracy.

In the future, we will continue our participation in the mock LISA data
challenges.  Challenge 3 has already been started and includes an SMBBH
data set where spin effects have been included in the waveform
\cite{mldc1b}. In order to participate, we must develop an analysis
technique to be able to search for inspiralling supermassive black
holes with spin.  Initially, we want to investigate how effectively we
are able to search for spinning binaries with non-spinning templates and
see if this approach might enable us to get a good estimate of the
masses and coalescence time of the binaries.  However, to obtain good
parameter estimates, we will need to incorporate the effects of spin
into our signal model.  Unfortunately the $\mathcal{F}$-statistic is not
directly applicable due to the added complications spinning binaries
bring.  We will either have to develop a new technique to analytically
maximize over some of the parameters, or be forced to place templates in
a much higher dimensional signal manifold.

\section*{Acknowledgements}

We would like to acknowledge many useful discussions with Curt Cutler,
Michele Vallisneri and Jeff Crowder.  This work has been suppored by an
SFTC grant (IWH and BSS) and the Royal Society (SF).

\bibliographystyle{unsrt}
\bibliography{references}

\end{document}